\documentclass[aps,prc,superscriptaddress,twocolumn,showpacs,nofootinbib]{revtex4-1}
\usepackage[utf8]{inputenc}
\usepackage{graphicx}   %       for graphics
\usepackage{latexsym}   %       for special symbols
\usepackage{enumerate}
\usepackage{rotating,booktabs,multirow}
\usepackage{amsmath}
\usepackage{amsfonts}
\usepackage{amssymb}
\usepackage{multirow}
\usepackage{bm}
\usepackage{xcolor}
\usepackage[colorlinks=true,linktocpage=true,linkcolor=blue,citecolor=blue,allcolors=blue]{hyperref}

\begin{document}
%============================================================================
\title{Decoding the flow evolution in Au+Au reactions at $1.23 A$ GeV \\ using hadron flow correlations and dileptons}
\author{Tom Reichert}
\affiliation{Institut f\"ur Theoretische Physik, Goethe Universit\"at Frankfurt, Max-von-Laue-Strasse 1, D-60438 Frankfurt am Main, Germany}
\affiliation{Helmholtz Research Academy Hesse for FAIR (HFHF), GSI Helmholtz Center for Heavy Ion Physics, Campus Frankfurt, Max-von-Laue-Str. 12, 60438 Frankfurt, Germany}

\author{Oleh Savchuk}
\affiliation{GSI Helmholtzzentrum f\"ur Schwerionenforschung GmbH, Planckstr. 1, 64291 Darmstadt, Germany}
\affiliation{Frankfurt Institute for Advanced Studies (FIAS), Ruth-Moufang-Str.1, D-60438 Frankfurt am Main, Germany}
\affiliation{Bogolyubov Institute for Theoretical Physics, 03680 Kyiv, Ukraine}
\affiliation{Facility for Rare Isotope Beams, Michigan State University, East Lansing, MI 48824 USA}

\author{Apiwit Kittiratpattana}
\affiliation{Institut f\"ur Theoretische Physik, Goethe Universit\"at Frankfurt, Max-von-Laue-Strasse 1, D-60438 Frankfurt am Main, Germany}
\affiliation{Suranaree University of Technology, University Avenue 111, Nakhon Ratchasima 30000, Thailand}

\author{Pengcheng~Li}
\affiliation{School of Nuclear Science and Technology, Lanzhou University, Lanzhou 730000, China}
\affiliation{School of Science, Huzhou University, Huzhou 313000, China}

\author{Jan~Steinheimer}
\affiliation{Frankfurt Institute for Advanced Studies (FIAS), Ruth-Moufang-Str.1, D-60438 Frankfurt am Main, Germany}

\author{Mark Gorenstein}
\affiliation{Frankfurt Institute for Advanced Studies (FIAS), Ruth-Moufang-Str.1, D-60438 Frankfurt am Main, Germany}
\affiliation{Bogolyubov Institute for Theoretical Physics, 03680 Kyiv, Ukraine}

\author{Marcus~Bleicher}
\affiliation{Institut f\"ur Theoretische Physik, Goethe Universit\"at Frankfurt, Max-von-Laue-Strasse 1, D-60438 Frankfurt am Main, Germany}
%\affiliation{GSI Helmholtzzentrum f\"ur Schwerionenforschung GmbH, Planckstr. 1, 64291 Darmstadt, Germany}
\affiliation{Helmholtz Research Academy Hesse for FAIR (HFHF), GSI Helmholtz Center for Heavy Ion Physics, Campus Frankfurt, Max-von-Laue-Str. 12, 60438 Frankfurt, Germany}

\begin{abstract}
We investigate the development of the directed, $v_1$, and elliptic flow, $v_2$, in heavy ion collisions in mid-central Au+Au reactions at $E_\mathrm{lab}=1.23 A$ GeV. We demonstrate that the elliptic flow of hot and dense matter is initially positive ($v_2>0$) due to the early pressure gradient. This positive $v_2$ transfers its momentum to the spectators, which leads to the creation of the directed flow $v_1$. In turn, the spectator shadowing of the in-plane expansion leads to a preferred decoupling of hadrons in the out-of-plane direction and results in a negative $v_2$ for the observable final state hadrons. We propose a measurement of $v_1-v_2$ flow correlations and of the elliptic flow of dileptons as  methods to pin down this evolution pattern. The elliptic flow of the dileptons allows then to determine the early-state EoS more precisely, because it avoids the strong modifications of the momentum distribution due to shadowing seen in the protons. This opens the unique opportunity for the HADES and CBM collaborations to measure the Equation-of-State directly at 2-3 times nuclear saturation density.
\end{abstract}

\maketitle

\section{Introduction}
Heavy-ion collisions provide an excellent opportunity to study nuclear matter in a controlled laboratory setting allowing to recreate conditions which were present in the early universe or are present in neutron stars. One of the main goals of nuclear research is the precise extraction of the nuclear Equation-of-State (EoS) at large baryon densities which are found in the interior of neutron stars. The link between astrophysical observations on a scale of $10^{15}$ m and collisions of heavy atomic nuclei at nearly the speed of light on a scale of $10^{-15}$ m comes as a surprise. Remarkable resemblance between binary neutron star mergers (BNS merger) and heavy ion collisions (HICs) has been found in Ref. \cite{Most:2022wgo}. Most intriguing is the fact that BNS mergers and HICs are subject to the same microscopic interactions, namely Quantum-Chromo-Dynamics (QCD). In order to better understand astrophysical objects, it is of utmost importance to pin down the nuclear Equation-of-State with high precision, desirably in a neutron rich system. 

HICs range from center of mass energies of 2 GeV at GSI's SIS18 accelerator up to several hundred GeV at BNL-RHIC or CERN-SPS or even to 13 TeV at CERN-LHC. However, the density range suited for BNS comparison will be probed in the upcoming FAIR facility which is currently constructed near Darmstadt, Germany. In this energy regime, the densities in central heavy ion collisions reaches from 2-6 times nuclear saturation density \cite{OmanaKuttan:2022the}. Currently, the lower range of this regime can already be probed in the currently running HADES experiment at GSI. Here, information about the Equation-of-State can be extracted via measurements of harmonic flow of hadrons \cite{Danielewicz:2002pu,LeFevre:2015paj,Reichert:2022gqe}, investigations of the speed of sound \cite{Bedaque:2014sqa,Tews:2018kmu,Steinheimer:2012bp} or strangeness enhancement or flow \cite{KAOS:2000ekm,Fuchs:2000kp,Hartnack:2005tr,HADES:2018noy,Chlad:2020kpm} and many more. Especially the second flow coefficient $v_2$ (called elliptic flow) is perfectly suited to investigate the EoS at large densities. It is well known from higher beam energies that the measured elliptic flow in the final state is directly connected to the pressure gradients exerted by the EoS. A detailed comparison of flow measurements at BNL's RHIC \cite{STAR:2000ekf,PHENIX:2003iij} with hydrodynamic simulations \cite{Huovinen:2001cy,Song:2007fn,Luzum:2008cw,Romatschke:2007mq} lead to the renowned finding that the deconfined phase of QCD (a Quark-Gluon-Plasma, QGP) resembles the most perfect liquid. 

At low energies the situation is more challenging: here the incoming baryon currents cannot decouple quickly enough to allow free expansion driven solely by the pressure gradients. Instead, the residues of the incoming nuclei (the spectators) are blocking the emission of particles in-plane during the compression phase. This leads to the well known squeeze-out effect of nucleons. Therefore, the final elliptic flow of protons is negative in the range $\sqrt{s_\mathrm{NN}}=2-4$~GeV, which is confirmed by measurements at FOPI \cite{FOPI:2004bfz,FOPI:2011aa}, EOS/E895 \cite{E895:1999ldn,E895:2000maf}, E877 \cite{E877:1997zjw}, STAR-BES \cite{STAR:2014clz,STAR:2020dav,STAR:2021ozh}.

Even though the magnitude of the negative elliptic flow is still sensitive to the EoS \cite{Steinheimer:2022gqb}, a detailed understanding of the dynamics which lead to this dependency and its connection to the directed flow is still missing. 

In this article we aim to understand the precise expansion dynamics during the compression stage and how the initial positive elliptic flow is observed as negative elliptic flow in the final state. This will allow to extract information on the Equation-of-State in the most dense stage of the reaction. We analyze the flow of Au-Au collisions at 1.23$A$ GeV kinetic beam energy, in line with recent HADES measurements \cite{HADES:2020lob,HADES:2022osk}, and investigate the contributions to the final observable flow. Finally we propose two distinct measurements (flow correlations and dilepton flow) to test our findings.

\section{Model setup and flow extraction}
For the present study we use the Ultra-relativistic Quantum Molecular Dynamics (UrQMD) model \cite{Bass:1998ca,Bleicher:1999xi,Bleicher:2022kcu} in its most recent version (v3.5). UrQMD is a microscopic transport simulation based on the explicit propagation of hadrons in phase-space. The imaginary part of the interactions is modeled via binary elastic and inelastic collisions, leading to resonance excitations and decays or color flux-tube formation and their fragmentation. The real part of the interaction potential is implemented via different equations of state, where in the present work a realistic chiral mean field EoS is used, see \cite{OmanaKuttan:2022the}. In its current version, UrQMD includes a large body of baryonic and mesonic resonances up to masses of 4~GeV. The model is well established in the GSI energy regime. For recent studies of the bulk dynamics, we refer the reader to \cite{Reichert:2020uxs,Reichert:2021ljd}. For the analysis of the integrated harmonic flows at SIS energies see \cite{Hillmann:2018nmd,Hillmann:2019wlt,Reichert:2022gqe,Reichert:2022yxq}. 

The flow coefficients are identified with the Fourier coefficients in the series expansion of the azimuthal angle distribution which can be written as
\begin{equation}
    \frac{\mathrm{d}N}{\mathrm{d}\phi} = 1 + 2\sum\limits_{n=1}^\infty v_n\cos(n(\phi-\Psi_{RP})),
\end{equation}
in which $v_n$ is n-th order flow coefficient, $\phi$ is the azimuthal angle and $\Psi_{RP}$ is the angle of the reaction plane. The HADES experiment uses a forward wall to reconstruct the event plane from the spectator nucleons \cite{Kardan:2017knj}. In the simulation, the spectator event plane is fixed and $\Psi_{RP}=0$ is used for the present analysis of the simulation. The flow coefficients are then calculated as
\begin{equation}
    v_n = \langle \cos(n(\phi-\Psi_{RP})) \rangle,
\end{equation}
where the average $\langle\cdot\rangle$ is taken over all nucleons in a fixed rapidity or transverse momentum range in a given event.

The reader should be aware that this definition is different than the one usually used at the highest beam energies where the flow components are defined with respect to specific event planes. The methodical difference has its origin in the way that flow develops at low beam energies. Here, the initial squeeze out of nucleons leads to a large negative elliptic flow component with respect to the reaction plane defined by the spectators. The initial compression phase is decisive for the observed flow and therefore the scaling of the initial eccentricity with final state flow known from large collision energies \cite{PHOBOS:2006fqf} is not as straightforward. Recently, scaling could be established in Refs. \cite{HADES:2022osk,Reichert:2022imz}, however with a negative sign. Thus, it can be expected that also the flow correlations will differ significantly from what was observed at high beam energies, which opens up new possibilities to study the equation of state responsible for the formation of flow.

\section{Results}
All results were obtained by simulating 20-30\% peripheral Au+Au collisions at $E_{\rm beam}=1.23A$~GeV kinetic beam energy with the UrQMD (v3.5) model. We employ the model with the Chiral Mean Field (CMF) Equation-of-State as it was shown to yield the best description of available data (cf. Ref. \cite{Steinheimer:2022gqb}). We note that in this energy regime, the results of the CMS-EoS are very similar to the ones obtained by a hard EoS \cite{Hillmann:2018nmd,Hillmann:2019wlt}. We focus our analysis on participating nucleons and also exclude nucleons that are bound in light clusters. It has been shown \cite{Hillmann:2019wlt} that both effects need to be taken into account to reliably describe the measured data on nucleon flow \cite{HADES:2020lob,HADES:2022osk}. The centrality is selected via impact parameter cuts following previous Monte-Carlo Glauber simulations \cite{HADES:2017def}.

\subsection{Flow development of the system}
In contrast to ultra-relativistic energies probed at RHIC and LHC where the initial baryon-currents decouple quickly allowing the overlap region to propagate its spatial anisotropy to the final state momentum space anisotropy \cite{PHOBOS:2006fqf}, low energy heavy-ion collisions are very different. Here, the baryon-currents of the initial impinging nuclei are present over the whole course of the compression and expansion phase, which results in a space and time dependent interplay between pressure gradients, spectator blocking, compression and expansion phase and corrections due to transport coefficients such as shear viscosity \cite{Reichert:2020oes} which influence the development of the finally observable flow. 

%%%%%%%%%%%%%%%%%%%%%%%%%%%%%%%%%%
\begin{figure}[t]
    \centering
    \includegraphics[width=\columnwidth]{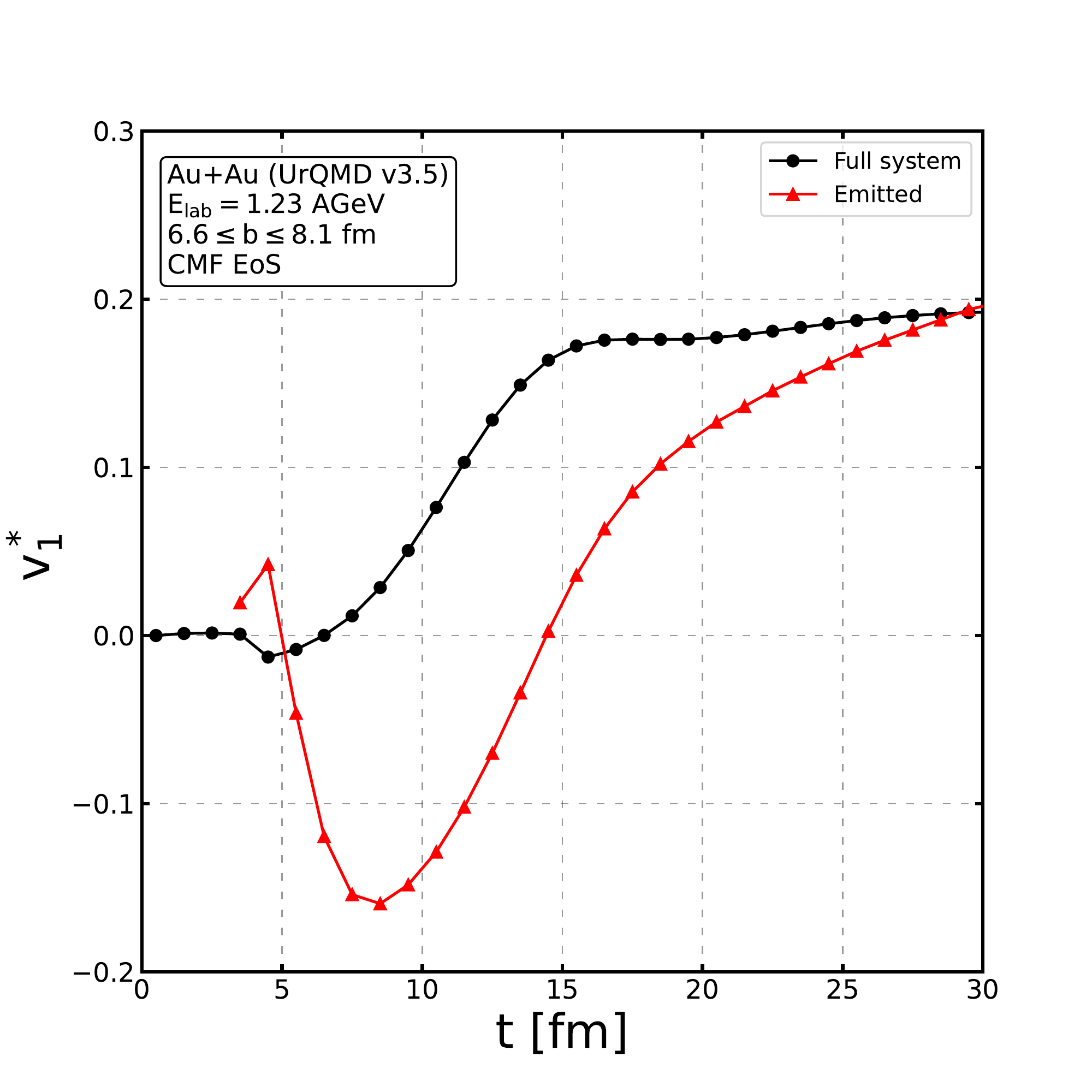}
    \caption{[Color online] The directed flow coefficient $v_1^*\equiv \left\langle {\rm sign}(y)\cdot v_1 \right\rangle$ calculated from all baryons (black circles) which are present in the whole phase space at time $t$ and of the nucleons emitted at time $t+\Delta t$ (red triangles) from 20-30\% peripheral Au+Au collisions at a kinetic beam energy of 1.23$A$ GeV from UrQMD.}
    \label{fig:v1_frz_and_timestep}
\end{figure}
%%%%%%%%%%%%%%%%%%%%%%%%%%%%%%%%%%
%%%%%%%%%%%%%%%%%%%%%%%%%%%%%%%%%%
\begin{figure}[t]
    \centering
    \includegraphics[width=\columnwidth]{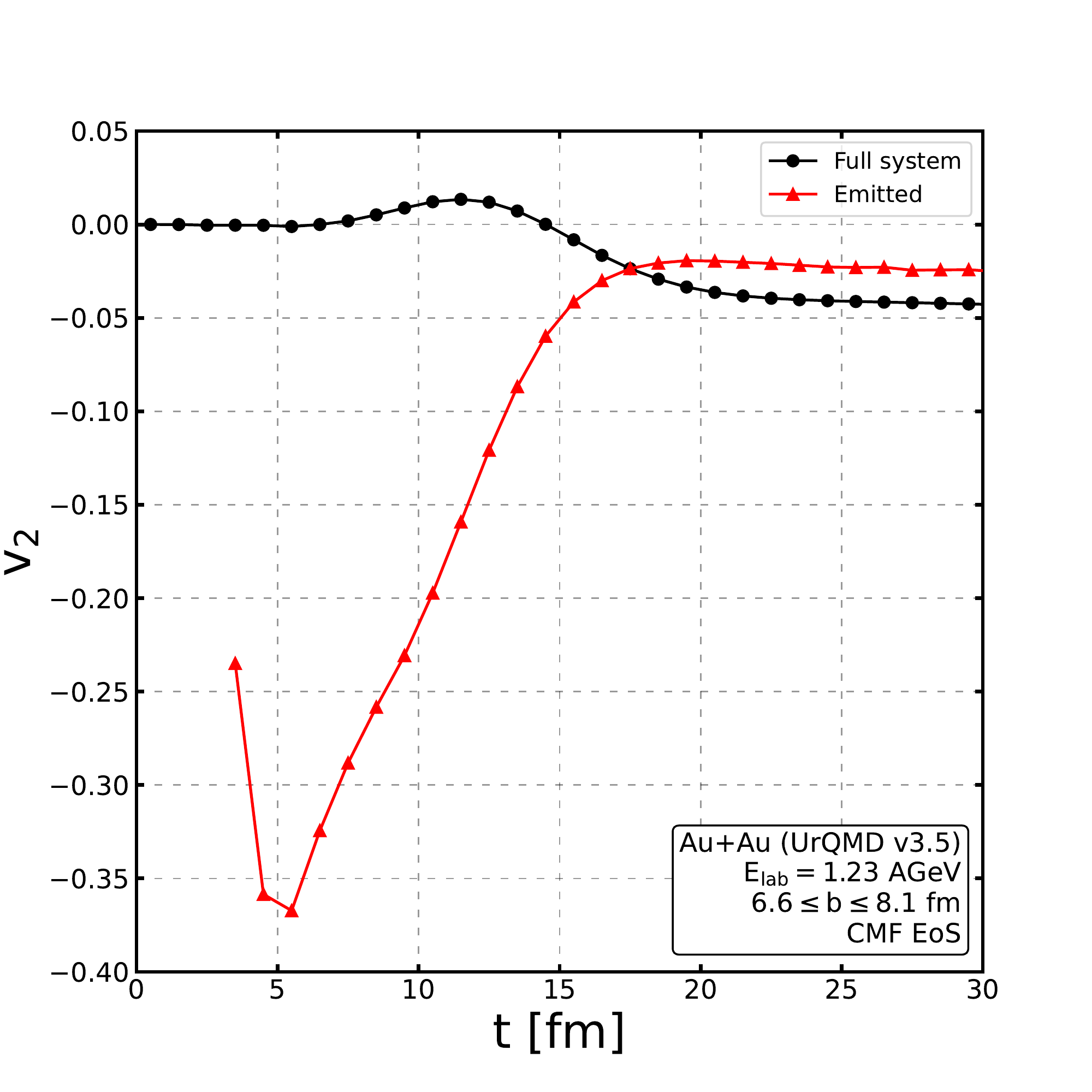}
    \caption{[Color online] The elliptic flow coefficient $v_2$ calculated from all baryons (black circles) which are present in the whole phase space at time $t$ and of the nucleons emitted at time $t$ (red triangles) from 20-30\% peripheral Au+Au collisions at a kinetic beam energy of 1.23$A$ GeV from UrQMD.}
    \label{fig:v2_frz_and_timestep}
\end{figure}
%%%%%%%%%%%%%%%%%%%%%%%%%%%%%%%%%%
%%%%%%%%%%%%%%%%%%%%%%%%%%%%%%%%%%

It is thus convenient to start the investigation by discussing how flow develops over the course of the collision in the whole system before we later quantify the different contributions to the final observable flow at freeze-out. For this we simulate semi-peripheral (corresponding to $6.6\leq b\leq8.1$~fm, cf. Ref. \cite{HADES:2017def}) Au+Au collisions with UrQMD. In this section we integrate over all baryons present at a specified time $t$ (the time is defined in the computational frame, i.e. the center-of-mass frame of the whole system) to extract the time development of the harmonic flow coefficients $v_1^*$ and $v_2$. We show the flow coefficients $v_1^*$ in Fig. \ref{fig:v1_frz_and_timestep} and $v_2$ in Fig. \ref{fig:v2_frz_and_timestep} calculated from all baryons which are present in the whole phase space at time $t$ as a black solid line with full circles. In addition we also show the directed and elliptic flow only of nucleons being emitted at given time $t+\Delta t$ as a red solid line with full triangles. It should be noted that, since the directed flow, $v_1$, averaged over all nucleons is exactly zero due to momentum conservation, we show the rapidity weighted averaged directed flow defined as
$v_1^*\equiv \left\langle {\rm sign}(y)\cdot v_1 \right\rangle$ where the average runs over all baryons and $y$ and $v_1$ correspond to the rapidity and directed flow of each baryon.

First, we observe that the time dependence of the directed flow calculated from all participating baryons present in the system (black line with circles) exerts a strong increase during the collision. During the initial compression stage the curve is consistent with zero until $\approx5-7$ fm/c when the density reaches 2-3 times saturation density and the CMF EoS becomes repulsive \cite{OmanaKuttan:2022the}. Here, maximum compression is reached \cite{OmanaKuttan:2022the}. At this time the development of the directed flow component begins due to a large pressure gradient in impact parameter direction. The pressure gradient pushes the nucleons in in-plane direction, which is reflected in a rapid development of large $v_1^*$ over the course of the next 7~fm. 

After 15~fm, the directed flow $v_1^*$ of all baryons present in the system reaches a plateau around $v_1^*\approx0.18$ in line with measured data \cite{HADES:2020lob,HADES:2022osk}. This tells that the bounce-off is generated in a rather short time frame and thus probes the system at maximal compression, i.e. $\rho_\mathrm{B}/\rho_0\approx2-3$ \cite{Reichert:2020oes,OmanaKuttan:2022the}.

The directed flow of the first nucleons that decouple kinetically from the system is strongly negative. This finding can be understood, because the nucleons which propagate in-plane will scatter with the residue of the projectile nucleus still propagating forward and therefore the only direction free for decoupling has a negative $v_1^*$ at initial contact. 

Turning to the elliptic flow $v_2$ (shown in Fig. \ref{fig:v2_frz_and_timestep}) supports this line of arguments. The elliptic flow of the whole system is zero until maximal compression is achieved at the time of full geometric overlap at 7~fm. This means that the system's momentum space remains isotropic in the transverse plane in momentum space from initial contact until full overlap. At overlap time the compression is maximal with values around $\rho_\mathrm{B}/\rho_0=2-3$ \cite{Reichert:2020oes,OmanaKuttan:2022the}. Although measurements by the HADES collaboration \cite{HADES:2020lob} show in agreement with theoretical models \cite{Hillmann:2019wlt,Mohs:2020awg} that the elliptic flow is negative in the final state (i.e. at kinetic freeze-out) at this energy, the elliptic flow of the whole system extracted at different times starts to become positive after full overlap is reached and only turns negative at even later stages of the evolution of the system. This can be interpreted as follows: At the time of maximal compression the overlap zone resembles an almond shape which is known also from high energies. The spatial pressure gradient of this shape exerts a force over a large surface in in-plane direction (i.e. aligned with the impact parameter). The participating baryons in the center of the system thus try to expand in-plane with positive $v_2$ but they cannot decouple because the spectators are still blocking\footnote{Here the phrase 'blocking' should not be understood as hard wall blocking, but as a deceleration or momentum transfer of the initially expanding system to the (semi-)spectator matter.} expansion in this direction. The resulting momentum transfer to the (semi-)spectators next to the central collision zone then generates the observed $v_1^*$ (as discussed above). However, the observed hadrons are emitted out-of-plane, because only in this direction the emission is not blocked and therefore the elliptic flow of the participating nucleons turns negative at a slightly later time around 15~fm which is exactly where the directed flow reaches its plateau. 

Contrary to the $v_2$ of all baryons in the system, the elliptic flow of the emitted nucleons is always strongly negative even though the full systems $v_2$ is zero or positive over the first 15~fm. In other words even though more nucleons are flowing in in-plane direction than out-of-plane, those in-plane are blocked from emission by the spectators and can not be observed as free hadrons in the detector. 

This underlines the importance of shadowing to understand flow at SIS energies: the pressure gradient generated by the Equation-of-State behaves similar to the highest beam energies and tries to generate expansion with positive $v_2$. However, the momentum flow in $x$-direction gets absorbed by nucleons close to the central reaction zone and the bypassing spectators generating a strongly positive $v_1^*$ (the bounce-off, or rather a ``push-away") and leads to an isotropization of the momentum flow in $x$-direction. In contrast, the nucleons which can freeze-out dominantly only with momenta in $y$-direction create a negative $v_2$. This also explains how a softening of the early EoS, causing less early expansion and smaller early $v_2$, leads to a smaller final $v_1^*$. 

\begin{figure} [t!hb]
    \centering
    \includegraphics[width=\columnwidth]{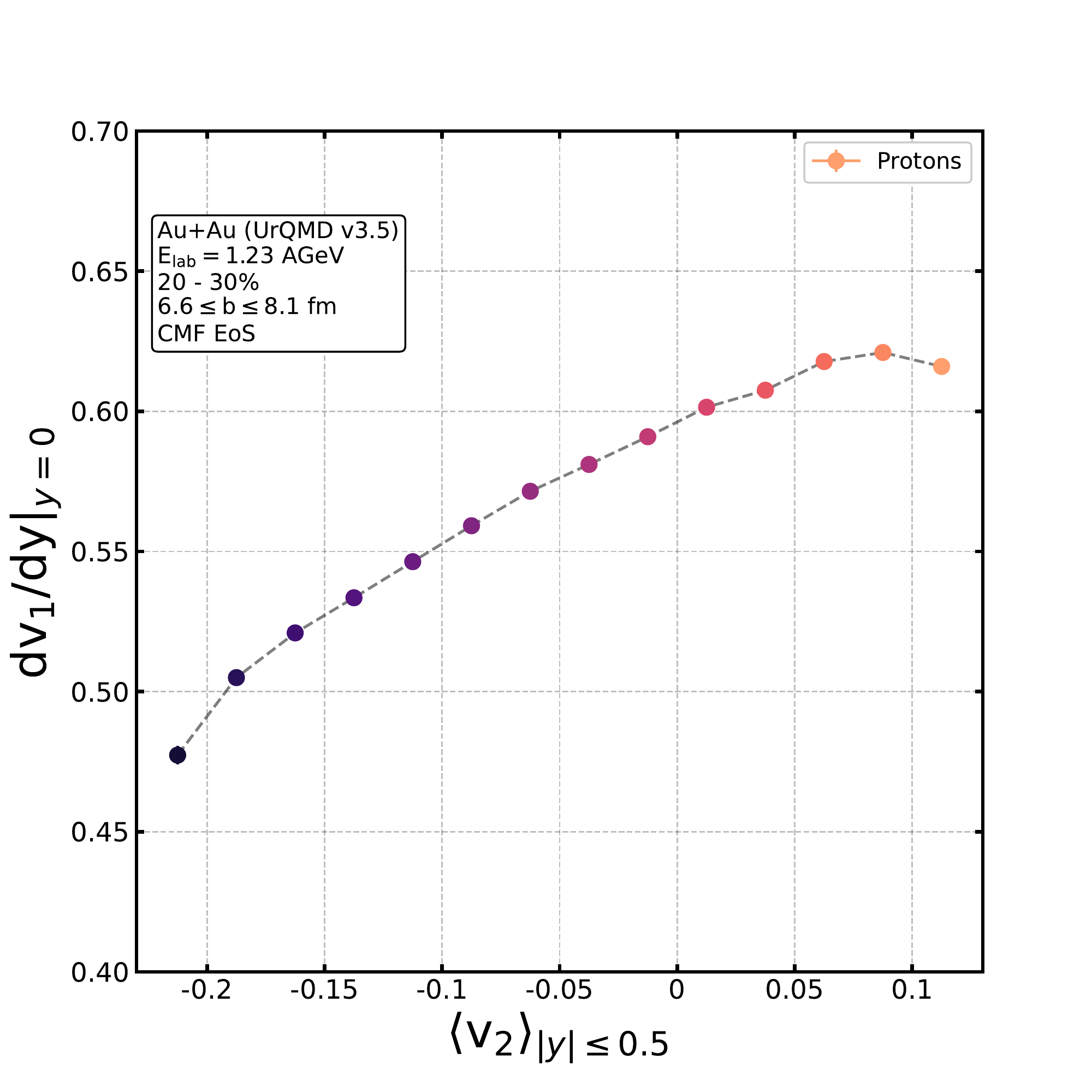}
    \caption{[Color online] The mid-rapidity slope of the directed flow of nucleons at kinetic freeze-out as a function of event-class selected integrated $v_2$ values at mid-rapidity from 20-30\% peripheral Au+Au collisions at a kinetic beam energy of 1.23$A$ GeV from UrQMD.}
    \label{fig:v1_v2_corr}
\end{figure}

The tight connection between $v_1^*$ and $v_2$ should therefore be observable in their correlation. Following our previous study in Ref. \cite{Reichert:2022gqe} we define event classes by triggering on the integrated final state elliptic flow of nucleons $\langle v_2\rangle_{|y|\leq0.5}$. As discussed, a smaller $v_2$ leads to less momentum transfer to the (semi-)spectators and thus the integrated final elliptic flow of protons should have a positive correlation with the mid-rapidity slope of the directed flow. Fig. \ref{fig:v1_v2_corr} shows the mid-rapidity slope of the directed flow of nucleons at kinetic freeze-out as a function of event-class selected integrated $v_2$ values at mid-rapidity from 20-30\% peripheral Au+Au collisions at a kinetic beam energy of 1.23$A$ GeV.
%%%%%%%%%%%%%%%%%%%%%%%%%%%%%%%%%%
%%%%%%%%%%%%%%%%%%%%%%%%%%%%%%%%%%
A clear positive correlation between $v_2$ and $\mathrm{d}v_1/\mathrm{d}y|_{y=0}$ is found. This result supports the idea that the initial expansion in $x$-direction exerts a momentum transfer to the (semi-spectator) nucleons close to the central reaction zone and thus enforcing a strong correlation between the initial elliptic flow and the final directed flow.

How can such a scenario be tested further? In the following section we suggest dileptons as a second observable to explore the early stage expansion pattern.

%%%%%%%%%%%%%%%%%%%%%%%%%%%%%%%%%%
\begin{figure} [t!hb]
    \centering
    \includegraphics[width=\columnwidth]{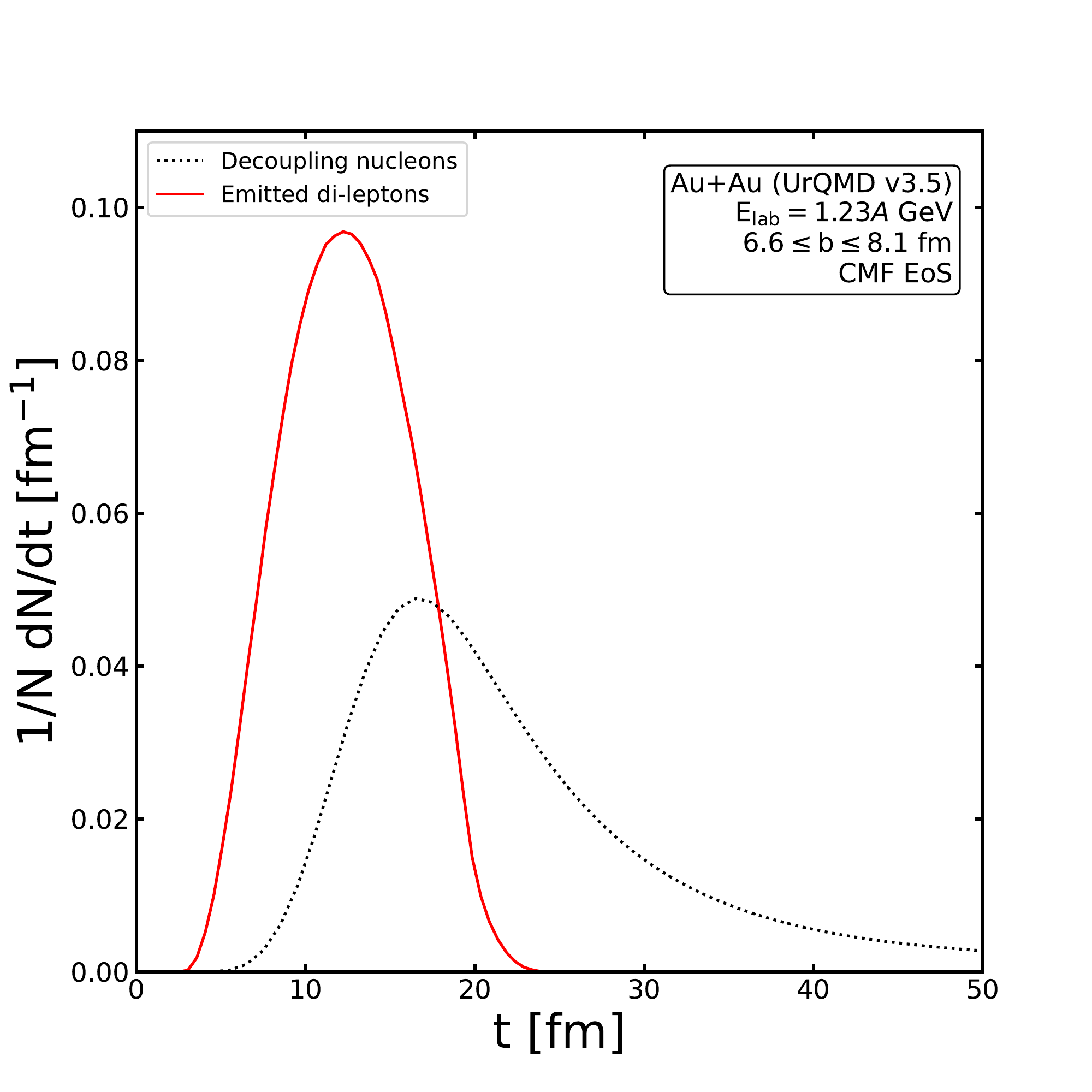}
    \caption{[Color online] The time distribution of nucleons at kinetic freeze-out (dotted black line) and of emitted dileptons (solid red line) from 20-30\% peripheral Au+Au collisions at a kinetic beam energy of 1.23 $A$GeV from UrQMD.}
    \label{fig:dNdt_frzout}
\end{figure}
%%%%%%%%%%%%%%%%%%%%%%%%%%%%%%%%%%

\subsection{Time of decoupling and measurement via dileptons}
In an experimental setup the time dependence of heavy-ion collisions cannot be accessed by hadronic observables. On the other hand, dilepton emission from the hot and dense phase offers time integrated information of the whole evolution history \cite{Shuryak:1978ij,McLerran:1984ay,Bratkovskaya:1996qe,Rapp:1999ej}. Dileptons are perfect candidates because they only interact electromagnetically and thus very weakly in comparison to the strong interaction \cite{Rapp:2014hha}. Recent dilepton measurements by the HADES collaboration \cite{HADES:2019auv} have shown great potential for future high precision measurements of the properties of the matter produced \cite{Savchuk:2022aev}.

To calculate the dilepton emission from our simulated collisions we will employ a well known coarse graining method \cite{Bravina:1998pi,Bravina:1999dh, Huovinen:2002im,Endres:2013daa,Endres:2014zua,Endres:2015fna,Endres:2015egk,Galatyuk:2015pkq,Endres:2016tkg,Reichert:2020yhx,Reichert:2020oes} where event averaged spatial distributions of the baryon and energy-momentum density are used to calculate the time dependent in-medium dilepton emission using state-of-the-art vector meson spectral functions \cite{Rapp:1999us,vanHees:2007th,Rapp:2013nxa}. In particular we run UrQMD-CMF events at a fixed impact parameter of $b=6.6$ fm and apply the coarse graining method described in \cite{Savchuk:2022aev}.

From the coarse grained simulation we can directly extract the time dependence of the dilepton emission rate, at mid-rapidity, and compare it to the time evolution and decoupling rate of the nucleons. In Fig. \ref{fig:dNdt_frzout} the normalized freeze-out time distributions of the nucleons are shown as a dotted black line while the emission time distribution of the dileptons is shown as a solid red line. One can clearly observe that the dilepton emission time distribution is much more narrow and centered at 10-12~fm. In contrast, the nucleon kinetic freeze-out time distribution is very broad, peaks at 18-20~fm and is strongly skewed towards later times. By comparison with the time development of the directed and elliptic flow shown previously in Figs. \ref{fig:v1_frz_and_timestep} and \ref{fig:v2_frz_and_timestep} one can see that the dileptons are emitted exactly during the most hot and dense phase which is the time when $v_1*$ is just generated and $v_2$ is positive. The nucleons, on the other hand, mainly decouple when the elliptic flow of the whole system has already turned negative due to immense shadowing and the directed flow has reached its plateau.

A similar idea has been put forward in Refs. \cite{Chatterjee:2005de,Chatterjee:2007xk,Chatterjee:2008tp,Deng:2010pq,vanHees:2011vb,Mohanty:2011fp,Vujanovic:2013jpa,Vovchenko:2016ijt} considering RHIC and LHC energies where the early $v_2$ of the partonic phase can be accessed via measurement of the flow of direct photons e.g. at PHENIX \cite{PHENIX:2011oxq} or at ALICE \cite{Lohner:2012ct,ALICE:2018dti}. 

We can now compare the elliptic flow of the dileptons with the hadrons at kinetic freeze-out. In particular the proton, $\pi$ and $\rho$(770) flows are  of special interest. The pions probe the kinetic decoupling stage while the $\rho$ either decays into a dilepton pair within the medium or into a $\pi\pi$ pair at kinetic freeze-out allowing us to study the final state and early flow in a mass range not dominated by pions. 

%%%%%%%%%%%%%%%%%%%%%%%%%%%%%%%%%%
\begin{figure}
    \centering
    \includegraphics[width=\columnwidth]{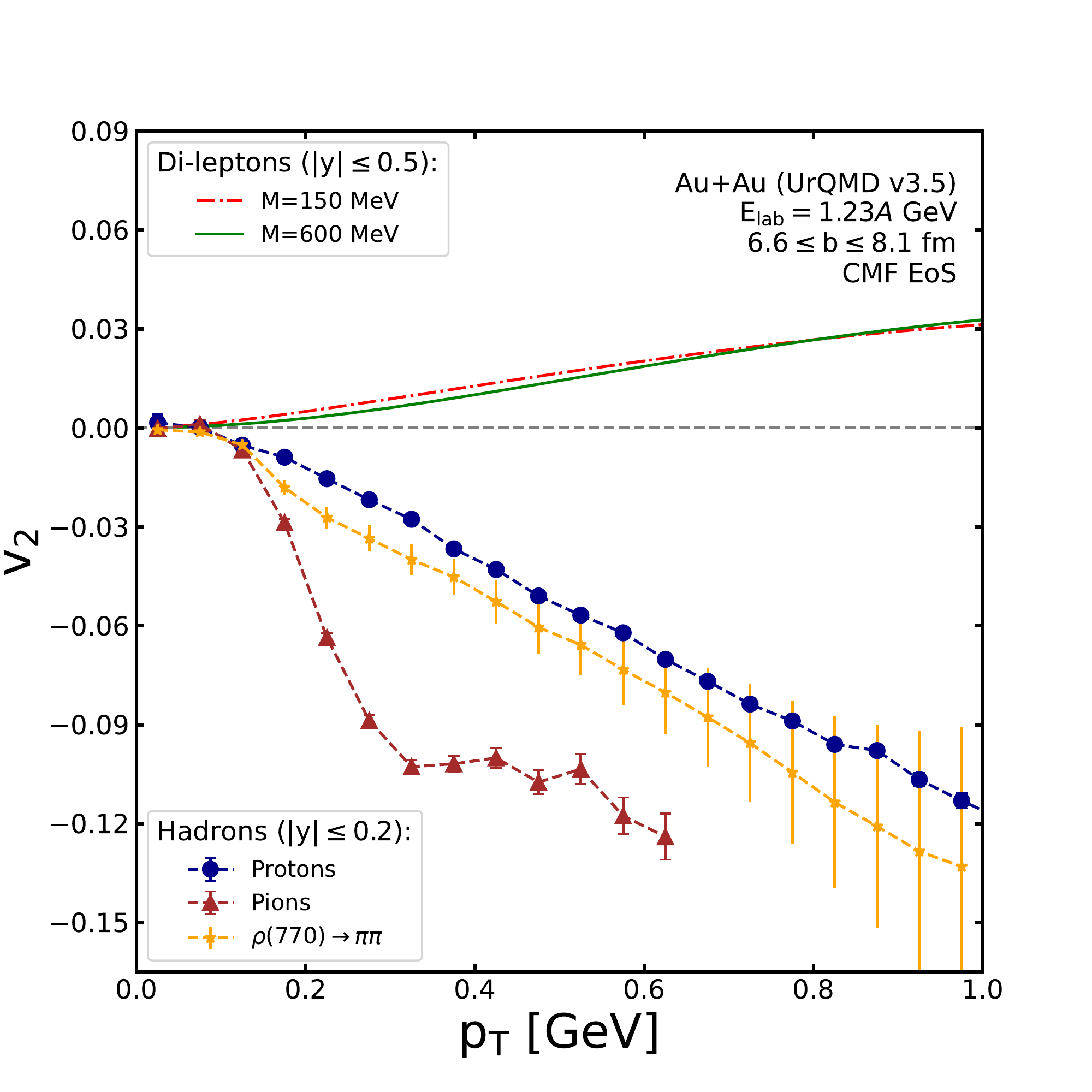}
    \caption{[Color online] The transverse momentum dependence of the elliptic flow of $\pi$ (brown triangles), protons (dark blue circles), final state $\rho$(770) (orange stars) and dileptons at invariant masses of $M=0.15$~GeV (red line) and $M=0.6$~GeV (green line) in peripheral Au+Au collisions at a kinetic beam energy of 1.23$A$ GeV from UrQMD.}
    \label{fig:v2_pt}
\end{figure}
%%%%%%%%%%%%%%%%%%%%%%%%%%%%%%%%%%

Figure \ref{fig:v2_pt} displays the transverse momentum dependence of the elliptic flow $v_2$ at mid-rapidity of the emitted dilepton radiation at invariant masses of $M=0.150$~GeV (dash-dotted red line) and $M=0.600$~GeV (green line) in comparison to the elliptic flow of protons (dark blue dashed line with circles), pions (brown dashed line with triangles) and final state $\rho$ mesons (dashed orange line with stars). 

As expected the elliptic flow of protons, pions and the $\rho$ mesons is negative and decreasing with increasing transverse momentum. (The calculations are also in agreement with the recent HADES measurements for protons, deuterons, tritons and preliminary pions at mid-rapidity \cite{HADES:2022osk}.) The dilepton $v_2$ on the other hand shows the opposite behavior: It probes the system during the most hot and dense phase when the pressure gradients generate expansion in in-plane direction and thence their elliptic flow is positive and increases with increasing transverse momentum. The dilepton flow at both invariant masses increases up to 3\% at 1 GeV transverse momentum which is in good alignment with the elliptic flow of the whole system (cf. Fig. \ref{fig:v2_frz_and_timestep}). One can further observe hydrodynamic flow scaling in the dileptons where larger masses are shifted to higher $p_\mathrm{T}$ \cite{Fries:2003kq}. 

We conclude that the above discussed effect of a positive elliptic flow experienced by all particles in the central collision zone exerted by the pressure gradients during the most dense phase can be seen in the flow of vector mesons decaying to dileptons throughout the time evolution. In contrast, stable hadrons and resonances that can be reconstructed in their hadronic decay channels are only sensitive to the last generation of particles at kinetic freeze-out and are therefore subject to a negative observable elliptic flow value. This allows to measure the nuclear Equation-of-State directly at 2-3 times saturation density using dileptons in the $\rho$-mass excess region.

\section{Discussion}
Before we conclude, we would like to address that similar time evolutions of the elliptic flow where presented in Refs. \cite{LeFevre:2016vpp,Fevre:2019xao,Gao:2022shr} within a QMD model at much lower energies, confirming the general expansion pattern of matter at such collision energies. However, the previous studies did not make any attempts to link the specific anisotropic expansion dynamics of the matter at the highest baryon densities to experimentally accessible observables. Here, we present for the first time two distinct observables that allow to pin down the expansion of the matter during the most dense stages, namely dilepton emission and flow correlations. Therefore, we suggest to scrutinize $v_1-v_2$ correlations and the positive elliptic flow of dileptons in the region of the $\rho$ mass as distinct signals for an initial expansion in positive $x$-direction, which is later turned by shadowing into a negative elliptic flow with an expansion predominantly along the $y$-direction.

\section{Conclusion}
We presented a detailed study of the time evolution of flow in Au-Au collisions at a beam energy of $1.23A$~GeV employing the Ultra-relativistic Quantum-Molecular-Dynamics model (UrQMD v3.5). The investigations revealed that, similar to known flow patterns at ultra-relativistic beam energies, the initial overlap geometry does lead to a positive elliptic flow during the early evolution of the system. This positive initial $v_2$ acts as the source of the directed flow, $v_1$. The negative $v_2$, observed in the final state hadrons, is mainly due to  shadowing effect. We suggest to use $v_1-v_2$ correlations to probe the described connection between the different flow components and to confirm the predicted initial state expansion geometry. We further suggest to use dileptons as another independent probe to observe the early expansion of the matter in $x$-direction, which will be signaled by a positive $v_2$ for dileptons in contrast to a negative $v_2$ (already observed) for protons.

As the early flow is a direct reaction of the system to the initial pressure gradient, and thus the Equation-of-State at the highest densities, the measurement of the dilepton elliptic flow may provide direct access to the high density EoS of QCD matter in a clearer way than possible before.

\begin{acknowledgements}
The authors thank Behruz Kardan and Christoph Blume as well as Arnaud Le F\`evre and J\"org Aichelin for fruitful discussion about ther previous work. This article is part of a project that has received funding from the European Union’s Horizon 2020 research and innovation programme under grant agreement STRONG – 2020 - No 824093. OS thanks the GSI for funding. Computational resources were provided by the Center for Scientific Computing (CSC) of the Goethe University and the ``Green Cube" at GSI, Darmstadt. This project was supported by the DAAD (PPP Thailand). This research has received funding support from the NSRF via the Program Management Unit for Human Resources \& Institutional Development, Research and Innovation [grant number B16F640076].
\end{acknowledgements}

%%%%%%%%%%%%%%%%%%%%%%%%%%%%%%%%%%%%%%%%%%%%%%%%%%%%%%%%%%%%%%%%%%%%%%%%%%%%%%%

%%%%%%%%%%%%%%%%%%%%%%%%%%%%%%%%%%%%%%%%%%%%%%%%%%%%%%%%%%%%%%%%%%%%%%%%%%%%%%% 


\begin{thebibliography}{100}
%%% 

%\cite{Most:2022wgo}
\bibitem{Most:2022wgo}
E.~R.~Most, A.~Motornenko, J.~Steinheimer, V.~Dexheimer, M.~Hanauske, L.~Rezzolla and H.~Stoecker,
%``Probing neutron-star matter in the lab: connecting binary mergers to heavy-ion collisions,''
[arXiv:2201.13150 [nucl-th]].
%19 citations counted in INSPIRE as of 17 Feb 2023

%\cite{OmanaKuttan:2022the}
\bibitem{OmanaKuttan:2022the}
M.~Omana Kuttan, A.~Motornenko, J.~Steinheimer, H.~Stoecker, Y.~Nara and M.~Bleicher,
%``A chiral mean-field equation-of-state in UrQMD: effects on the heavy ion compression stage,''
Eur. Phys. J. C \textbf{82}, no.5, 427 (2022)
doi:10.1140/epjc/s10052-022-10400-2
[arXiv:2201.01622 [nucl-th]].
%12 citations counted in INSPIRE as of 17 Feb 2023

%\cite{Danielewicz:2002pu}
\bibitem{Danielewicz:2002pu}
P.~Danielewicz, R.~Lacey and W.~G.~Lynch,
%``Determination of the equation of state of dense matter,''
Science \textbf{298}, 1592-1596 (2002)
doi:10.1126/science.1078070
[arXiv:nucl-th/0208016 [nucl-th]].
%1163 citations counted in INSPIRE as of 23 Feb 2023

%\cite{LeFevre:2015paj}
\bibitem{LeFevre:2015paj}
A.~Le F\`evre, Y.~Leifels, W.~Reisdorf, J.~Aichelin and C.~Hartnack,
%``Constraining the nuclear matter equation of state around twice saturation density,''
Nucl. Phys. A \textbf{945}, 112-133 (2016)
doi:10.1016/j.nuclphysa.2015.09.015
[arXiv:1501.05246 [nucl-ex]].
%103 citations counted in INSPIRE as of 21 Feb 2023

%\cite{Reichert:2022gqe}
\bibitem{Reichert:2022gqe}
T.~Reichert, J.~Steinheimer, C.~Herold, A.~Limphirat and M.~Bleicher,
%``Harmonic flow correlations in Au+Au reactions at 1.23 AGeV: a new testing ground for the equation-of-state and expansion geometry,''
Eur. Phys. J. C \textbf{82}, no.6, 510 (2022)
doi:10.1140/epjc/s10052-022-10480-0
[arXiv:2203.15550 [nucl-th]].
%4 citations counted in INSPIRE as of 25 Jan 2023

%\cite{Bedaque:2014sqa}
\bibitem{Bedaque:2014sqa}
P.~Bedaque and A.~W.~Steiner,
%``Sound velocity bound and neutron stars,''
Phys. Rev. Lett. \textbf{114}, no.3, 031103 (2015)
doi:10.1103/PhysRevLett.114.031103
[arXiv:1408.5116 [nucl-th]].
%204 citations counted in INSPIRE as of 10 Feb 2023

%\cite{Tews:2018kmu}
\bibitem{Tews:2018kmu}
I.~Tews, J.~Carlson, S.~Gandolfi and S.~Reddy,
%``Constraining the speed of sound inside neutron stars with chiral effective field theory interactions and observations,''
Astrophys. J. \textbf{860}, no.2, 149 (2018)
doi:10.3847/1538-4357/aac267
[arXiv:1801.01923 [nucl-th]].
%233 citations counted in INSPIRE as of 10 Feb 2023

%\cite{Steinheimer:2012bp}
\bibitem{Steinheimer:2012bp}
J.~Steinheimer and M.~Bleicher,
%``Extraction of the sound velocity from rapidity spectra: Evidence for QGP formation at FAIR/RHIC-BES energies,''
Eur. Phys. J. A \textbf{48}, 100 (2012)
doi:10.1140/epja/i2012-12100-0
[arXiv:1207.2792 [nucl-th]].
%19 citations counted in INSPIRE as of 08 Dec 2022

%\cite{KAOS:2000ekm}
\bibitem{KAOS:2000ekm}
C.~T.~Sturm \textit{et al.} [KAOS],
%``Evidence for a soft nuclear equation of state from kaon production in heavy ion collisions,''
Phys. Rev. Lett. \textbf{86}, 39-42 (2001)
doi:10.1103/PhysRevLett.86.39
[arXiv:nucl-ex/0011001 [nucl-ex]].
%263 citations counted in INSPIRE as of 21 Feb 2023

%\cite{Fuchs:2000kp}
\bibitem{Fuchs:2000kp}
C.~Fuchs, A.~Faessler, E.~Zabrodin and Y.~M.~Zheng,
%``Probing the nuclear equation of state by K+ production in heavy ion collisions,''
Phys. Rev. Lett. \textbf{86}, 1974-1977 (2001)
doi:10.1103/PhysRevLett.86.1974
[arXiv:nucl-th/0011102 [nucl-th]].
%227 citations counted in INSPIRE as of 21 Feb 2023

%\cite{Hartnack:2005tr}
\bibitem{Hartnack:2005tr}
C.~Hartnack, H.~Oeschler and J.~Aichelin,
%``Hadronic matter is soft,''
Phys. Rev. Lett. \textbf{96}, 012302 (2006)
doi:10.1103/PhysRevLett.96.012302
[arXiv:nucl-th/0506087 [nucl-th]].
%161 citations counted in INSPIRE as of 21 Feb 2023

%\cite{HADES:2018noy}
\bibitem{HADES:2018noy}
J.~Adamczewski-Musch \textit{et al.} [HADES],
%``Sub-threshold production of K$^{0}_{s}$ mesons and ${\Lambda}$ hyperons in Au+Au collisions at $\sqrt{s_{NN}}$ = 2.4 GeV,''
Phys. Lett. B \textbf{793}, 457-463 (2019)
doi:10.1016/j.physletb.2019.03.065
[arXiv:1812.07304 [nucl-ex]].
%37 citations counted in INSPIRE as of 27 Feb 2023

%\cite{Chlad:2020kpm}
\bibitem{Chlad:2020kpm}
L.~Chlad [HADES],
%``Strangeness Flow in $\mathrm {Au}+\mathrm {Au}$ Collisions at $1.23\,\mathrm {AGeV}$ Measured with HADES,''
Springer Proc. Phys. \textbf{250}, 221-224 (2020)
doi:10.1007/978-3-030-53448-6\_33
%1 citations counted in INSPIRE as of 29 Nov 2022

%\cite{STAR:2000ekf}
\bibitem{STAR:2000ekf}
K.~H.~Ackermann \textit{et al.} [STAR],
%``Elliptic flow in Au + Au collisions at (S(NN))**(1/2) = 130 GeV,''
Phys. Rev. Lett. \textbf{86}, 402-407 (2001)
doi:10.1103/PhysRevLett.86.402
[arXiv:nucl-ex/0009011 [nucl-ex]].
%813 citations counted in INSPIRE as of 27 Feb 2023

%\cite{PHENIX:2003iij}
\bibitem{PHENIX:2003iij}
S.~S.~Adler \textit{et al.} [PHENIX],
%``Identified charged particle spectra and yields in Au+Au collisions at S(NN)**1/2 = 200-GeV,''
Phys. Rev. C \textbf{69}, 034909 (2004)
doi:10.1103/PhysRevC.69.034909
[arXiv:nucl-ex/0307022 [nucl-ex]].
%983 citations counted in INSPIRE as of 31 Jan 2023

%\cite{Huovinen:2001cy}
\bibitem{Huovinen:2001cy}
P.~Huovinen, P.~F.~Kolb, U.~W.~Heinz, P.~V.~Ruuskanen and S.~A.~Voloshin,
%``Radial and elliptic flow at RHIC: Further predictions,''
Phys. Lett. B \textbf{503}, 58-64 (2001)
doi:10.1016/S0370-2693(01)00219-2
[arXiv:hep-ph/0101136 [hep-ph]].
%975 citations counted in INSPIRE as of 27 Feb 2023

%\cite{Song:2007fn}
\bibitem{Song:2007fn}
H.~Song and U.~W.~Heinz,
%``Suppression of elliptic flow in a minimally viscous quark-gluon plasma,''
Phys. Lett. B \textbf{658}, 279-283 (2008)
doi:10.1016/j.physletb.2007.11.019
[arXiv:0709.0742 [nucl-th]].
%398 citations counted in INSPIRE as of 09 Feb 2023

%\cite{Luzum:2008cw}
\bibitem{Luzum:2008cw}
M.~Luzum and P.~Romatschke,
%``Conformal Relativistic Viscous Hydrodynamics: Applications to RHIC results at s(NN)**(1/2) = 200-GeV,''
Phys. Rev. C \textbf{78}, 034915 (2008)
[erratum: Phys. Rev. C \textbf{79}, 039903 (2009)]
doi:10.1103/PhysRevC.78.034915
[arXiv:0804.4015 [nucl-th]].
%896 citations counted in INSPIRE as of 27 Feb 2023

%\cite{Romatschke:2007mq}
\bibitem{Romatschke:2007mq}
P.~Romatschke and U.~Romatschke,
%``Viscosity Information from Relativistic Nuclear Collisions: How Perfect is the Fluid Observed at RHIC?,''
Phys. Rev. Lett. \textbf{99}, 172301 (2007)
doi:10.1103/PhysRevLett.99.172301
[arXiv:0706.1522 [nucl-th]].
%1090 citations counted in INSPIRE as of 27 Feb 2023

%\cite{FOPI:2004bfz}
\bibitem{FOPI:2004bfz}
A.~Andronic \textit{et al.} [FOPI],
%``Excitation function of elliptic flow in Au+Au collisions and the nuclear matter equation of state,''
Phys. Lett. B \textbf{612}, 173-180 (2005)
doi:10.1016/j.physletb.2005.02.060
[arXiv:nucl-ex/0411024 [nucl-ex]].
%182 citations counted in INSPIRE as of 27 Feb 2023

%\cite{FOPI:2011aa}
\bibitem{FOPI:2011aa}
W.~Reisdorf \textit{et al.} [FOPI],
%``Systematics of azimuthal asymmetries in heavy ion collisions in the 1 A GeV regime,''
Nucl. Phys. A \textbf{876}, 1-60 (2012)
doi:10.1016/j.nuclphysa.2011.12.006
[arXiv:1112.3180 [nucl-ex]].
%121 citations counted in INSPIRE as of 16 Feb 2023

%\cite{E895:1999ldn}
\bibitem{E895:1999ldn}
C.~Pinkenburg \textit{et al.} [E895],
%``Elliptic flow: Transition from out-of-plane to in-plane emission in Au + Au collisions,''
Phys. Rev. Lett. \textbf{83}, 1295-1298 (1999)
doi:10.1103/PhysRevLett.83.1295
[arXiv:nucl-ex/9903010 [nucl-ex]].
%225 citations counted in INSPIRE as of 03 Feb 2023

%\cite{E895:2000maf}
\bibitem{E895:2000maf}
H.~Liu \textit{et al.} [E895],
%``Sideward flow in Au + Au collisions between 2-A-GeV and 8-A-GeV,''
Phys. Rev. Lett. \textbf{84}, 5488-5492 (2000)
doi:10.1103/PhysRevLett.84.5488
[arXiv:nucl-ex/0005005 [nucl-ex]].
%111 citations counted in INSPIRE as of 24 Feb 2023

%\cite{E877:1997zjw}
\bibitem{E877:1997zjw}
J.~Barrette \textit{et al.} [E877],
%``Proton and pion production relative to the reaction plane in Au + Au collisions at AGS energies,''
Phys. Rev. C \textbf{56}, 3254-3264 (1997)
doi:10.1103/PhysRevC.56.3254
[arXiv:nucl-ex/9707002 [nucl-ex]].
%177 citations counted in INSPIRE as of 01 Feb 2023

%\cite{STAR:2014clz}
\bibitem{STAR:2014clz}
L.~Adamczyk \textit{et al.} [STAR],
%``Beam-Energy Dependence of the Directed Flow of Protons, Antiprotons, and Pions in Au+Au Collisions,''
Phys. Rev. Lett. \textbf{112}, no.16, 162301 (2014)
doi:10.1103/PhysRevLett.112.162301
[arXiv:1401.3043 [nucl-ex]].
%254 citations counted in INSPIRE as of 27 Feb 2023

%\cite{STAR:2020dav}
\bibitem{STAR:2020dav}
J.~Adam \textit{et al.} [STAR],
%``Flow and interferometry results from Au+Au collisions at $\sqrt{s_{NN}} = 4.5$ GeV,''
Phys. Rev. C \textbf{103}, no.3, 034908 (2021)
doi:10.1103/PhysRevC.103.034908
[arXiv:2007.14005 [nucl-ex]].
%52 citations counted in INSPIRE as of 27 Feb 2023

%\cite{STAR:2021ozh}
\bibitem{STAR:2021ozh}
M.~S.~Abdallah \textit{et al.} [STAR],
%``Light nuclei collectivity from~$\sqrt{s_{NN}}$ = 3 GeV Au+Au collisions at RHIC,''
Phys. Lett. B \textbf{827}, 136941 (2022)
doi:10.1016/j.physletb.2022.136941
[arXiv:2112.04066 [nucl-ex]].
%14 citations counted in INSPIRE as of 24 Feb 2023

%\cite{Steinheimer:2022gqb}
\bibitem{Steinheimer:2022gqb}
J.~Steinheimer, A.~Motornenko, A.~Sorensen, Y.~Nara, V.~Koch and M.~Bleicher,
%``The high-density equation of state in heavy-ion collisions: constraints from proton flow,''
Eur. Phys. J. C \textbf{82}, no.10, 911 (2022)
doi:10.1140/epjc/s10052-022-10894-w
[arXiv:2208.12091 [nucl-th]].
%13 citations counted in INSPIRE as of 24 Feb 2023

%\cite{HADES:2020lob}
\bibitem{HADES:2020lob}
J.~Adamczewski-Musch \textit{et al.} [HADES],
%``Directed, Elliptic, and Higher Order Flow Harmonics of Protons, Deuterons, and Tritons in $\mathrm{Au}+\mathrm{Au}$ Collisions at $\sqrt{{s}_{NN}}=2.4\text{ }\text{ }\mathrm{GeV}$,''
Phys. Rev. Lett. \textbf{125}, 262301 (2020)
doi:10.1103/PhysRevLett.125.262301
[arXiv:2005.12217 [nucl-ex]].
%40 citations counted in INSPIRE as of 27 Feb 2023

%\cite{HADES:2022osk}
\bibitem{HADES:2022osk}
J.~Adamczewski-Musch \textit{et al.} [HADES],
%``Proton, deuteron and triton flow measurements in Au+Au collisions at $\sqrt{s_{NN}} = 2.4$ GeV,''
[arXiv:2208.02740 [nucl-ex]].
%4 citations counted in INSPIRE as of 21 Feb 2023

%\cite{Bass:1998ca}
\bibitem{Bass:1998ca}
S.~A.~Bass, M.~Belkacem, M.~Bleicher, M.~Brandstetter, L.~Bravina, C.~Ernst, L.~Gerland, M.~Hofmann, S.~Hofmann and J.~Konopka, \textit{et al.}
%``Microscopic models for ultrarelativistic heavy ion collisions,''
Prog. Part. Nucl. Phys. \textbf{41}, 255-369 (1998)
doi:10.1016/S0146-6410(98)00058-1
[arXiv:nucl-th/9803035 [nucl-th]].
%1862 citations counted in INSPIRE as of 27 Feb 2023

%\cite{Bleicher:1999xi}
\bibitem{Bleicher:1999xi}
M.~Bleicher, E.~Zabrodin, C.~Spieles, S.~A.~Bass, C.~Ernst, S.~Soff, L.~Bravina, M.~Belkacem, H.~Weber and H.~Stoecker, \textit{et al.}
%``Relativistic hadron hadron collisions in the ultrarelativistic quantum molecular dynamics model,''
J. Phys. G \textbf{25}, 1859-1896 (1999)
doi:10.1088/0954-3899/25/9/308
[arXiv:hep-ph/9909407 [hep-ph]].
%1528 citations counted in INSPIRE as of 27 Feb 2023

%\cite{Bleicher:2022kcu}
\bibitem{Bleicher:2022kcu}
M.~Bleicher and E.~Bratkovskaya,
%``Modelling relativistic heavy-ion collisions with dynamical transport approaches,''
Prog. Part. Nucl. Phys. \textbf{122}, 103920 (2022)
doi:10.1016/j.ppnp.2021.103920
%18 citations counted in INSPIRE as of 21 Feb 2023

%\cite{Reichert:2020uxs}
\bibitem{Reichert:2020uxs}
T.~Reichert, P.~Hillmann and M.~Bleicher,
%``$\Delta$ resonances in Ca+Ca, Ni+Ni and Au+Au reactions from 1 AGeV to 2 AGeV: Consistency between yields, mass shifts and decoupling temperatures,''
Nucl. Phys. A \textbf{1007}, 122058 (2021)
doi:10.1016/j.nuclphysa.2020.122058
[arXiv:2004.10539 [nucl-th]].
%10 citations counted in INSPIRE as of 07 Feb 2023

%\cite{Reichert:2021ljd}
\bibitem{Reichert:2021ljd}
T.~Reichert, A.~Elz, T.~Song, G.~Coci, M.~Winn, E.~Bratkovskaya, J.~Aichelin, J.~Steinheimer and M.~Bleicher,
%``Comparison of heavy ion transport simulations: Ag + Ag collisions at $E_{lab}$ = 1.58A GeV,''
J. Phys. G \textbf{49}, no.5, 055108 (2022)
doi:10.1088/1361-6471/ac5dfe
[arXiv:2111.07652 [nucl-th]].
%11 citations counted in INSPIRE as of 21 Feb 2023

%\cite{Hillmann:2018nmd}
\bibitem{Hillmann:2018nmd}
P.~Hillmann, J.~Steinheimer and M.~Bleicher,
%``Directed, elliptic and triangular flow of protons in Au+Au reactions at 1.23 A GeV: a theoretical analysis of the recent HADES data,''
J. Phys. G \textbf{45}, no.8, 085101 (2018)
doi:10.1088/1361-6471/aac96f
[arXiv:1802.01951 [nucl-th]].
%35 citations counted in INSPIRE as of 25 Jan 2023

%\cite{Hillmann:2019wlt}
\bibitem{Hillmann:2019wlt}
P.~Hillmann, J.~Steinheimer, T.~Reichert, V.~Gaebel, M.~Bleicher, S.~Sombun, C.~Herold and A.~Limphirat,
%``First, second, third and fourth flow harmonics of deuterons and protons in Au+Au reactions at 1.23 AGeV,''
J. Phys. G \textbf{47}, no.5, 055101 (2020)
doi:10.1088/1361-6471/ab6fcf
[arXiv:1907.04571 [nucl-th]].
%33 citations counted in INSPIRE as of 15 Feb 2023

%\cite{Reichert:2022yxq}
\bibitem{Reichert:2022yxq}
T.~Reichert, J.~Steinheimer and M.~Bleicher,
%``3-dimensional flow analysis: A novel tool to study the collision geometry and the Equation-of-State,''
[arXiv:2207.02594 [nucl-th]].
%1 citations counted in INSPIRE as of 17 Feb 2023

%\cite{Kardan:2017knj}
\bibitem{Kardan:2017knj}
B.~Kardan [HADES],
%``Collective flow measurements with HADES in Au+Au collisions at 1.23A GeV,''
Nucl. Phys. A \textbf{967}, 812-815 (2017)
doi:10.1016/j.nuclphysa.2017.05.026
%12 citations counted in INSPIRE as of 27 Nov 2022

%\cite{PHOBOS:2006fqf}
\bibitem{PHOBOS:2006fqf}
C.~Loizides \textit{et al.} [PHOBOS],
%``Elliptic flow, eccentricity and eccentricity fluctuations,''
Braz. J. Phys. \textbf{37}, 770-772 (2007)
[arXiv:nucl-ex/0611017 [nucl-ex]].
%3 citations counted in INSPIRE as of 25 Jan 2023

%\cite{Reichert:2022imz}
\bibitem{Reichert:2022imz}
T.~Reichert, A.~Kittiratpattana, P.~Li, J.~Steinheimer and M.~Bleicher,
%``A Systematic Comparison of Ag+Ag and Au+Au Reactions at 1.23 AGeV,''
[arXiv:2208.10871 [nucl-th]].
%0 citations counted in INSPIRE as of 12 Jan 2023

%\cite{HADES:2017def}
\bibitem{HADES:2017def}
J.~Adamczewski-Musch \textit{et al.} [HADES],
%``Centrality determination of Au + Au collisions at 1.23A GeV with HADES,''
Eur. Phys. J. A \textbf{54}, no.5, 85 (2018)
doi:10.1140/epja/i2018-12513-7
[arXiv:1712.07993 [nucl-ex]].
%51 citations counted in INSPIRE as of 27 Feb 2023

%\cite{Reichert:2020oes}
\bibitem{Reichert:2020oes}
T.~Reichert, G.~Inghirami and M.~Bleicher,
%``A first estimate of $\eta/s$ in Au+Au reactions at $E_{lab}$ = 1.23 A GeV,''
Phys. Lett. B \textbf{817}, 136285 (2021)
doi:10.1016/j.physletb.2021.136285
[arXiv:2011.04546 [nucl-th]].
%12 citations counted in INSPIRE as of 20 Jan 2023

%\cite{Mohs:2020awg}
\bibitem{Mohs:2020awg}
J.~Mohs, M.~Ege, H.~Elfner and M.~Mayer,
%``Collective flow at SIS energies within a hadronic transport approach: Influence of light nuclei formation and equation~of state,''
Phys. Rev. C \textbf{105}, no.3, 034906 (2022)
doi:10.1103/PhysRevC.105.034906
[arXiv:2012.11454 [nucl-th]].
%15 citations counted in INSPIRE as of 21 Feb 2023

%\cite{Shuryak:1978ij}
\bibitem{Shuryak:1978ij}
E.~V.~Shuryak,
%``Quark-Gluon Plasma and Hadronic Production of Leptons, Photons and Psions,''
Phys. Lett. B \textbf{78}, 150 (1978)
doi:10.1016/0370-2693(78)90370-2
%865 citations counted in INSPIRE as of 27 Feb 2023

%\cite{McLerran:1984ay}
\bibitem{McLerran:1984ay}
L.~D.~McLerran and T.~Toimela,
%``Photon and Dilepton Emission from the Quark - Gluon Plasma: Some General Considerations,''
Phys. Rev. D \textbf{31}, 545 (1985)
doi:10.1103/PhysRevD.31.545
%564 citations counted in INSPIRE as of 20 Feb 2023

%\cite{Bratkovskaya:1996qe}
\bibitem{Bratkovskaya:1996qe}
E.~L.~Bratkovskaya and W.~Cassing,
%``Dilepton production from AGS to SPS energies within a relativistic transport approach,''
Nucl. Phys. A \textbf{619}, 413-446 (1997)
doi:10.1016/S0375-9474(97)00140-1
[arXiv:nucl-th/9611042 [nucl-th]].
%120 citations counted in INSPIRE as of 03 Dec 2022

%\cite{Rapp:1999ej}
\bibitem{Rapp:1999ej}
R.~Rapp and J.~Wambach,
%``Chiral symmetry restoration and dileptons in relativistic heavy ion collisions,''
Adv. Nucl. Phys. \textbf{25}, 1 (2000)
doi:10.1007/0-306-47101-9\_1
[arXiv:hep-ph/9909229 [hep-ph]].
%868 citations counted in INSPIRE as of 24 Feb 2023

%\cite{Rapp:2014hha}
\bibitem{Rapp:2014hha}
R.~Rapp and H.~van Hees,
%``Thermal Dileptons as Fireball Thermometer and Chronometer,''
Phys. Lett. B \textbf{753}, 586-590 (2016)
doi:10.1016/j.physletb.2015.12.065
[arXiv:1411.4612 [hep-ph]].
%112 citations counted in INSPIRE as of 21 Feb 2023

%\cite{HADES:2019auv}
\bibitem{HADES:2019auv}
J.~Adamczewski-Musch \textit{et al.} [HADES],
%``Probing dense baryon-rich matter with virtual photons,''
Nature Phys. \textbf{15}, no.10, 1040-1045 (2019)
doi:10.1038/s41567-019-0583-8
%93 citations counted in INSPIRE as of 21 Feb 2023

%\cite{Savchuk:2022aev}
\bibitem{Savchuk:2022aev}
O.~Savchuk, A.~Motornenko, J.~Steinheimer, V.~Vovchenko, M.~Bleicher, M.~Gorenstein and T.~Galatyuk,
%``Enhanced dilepton emission from a phase transition in dense matter,''
[arXiv:2209.05267 [nucl-th]].
%3 citations counted in INSPIRE as of 17 Feb 2023

%\cite{Bravina:1998pi}
\bibitem{Bravina:1998pi}
L.~V.~Bravina, M.~I.~Gorenstein, M.~Belkacem, S.~A.~Bass, M.~Bleicher, M.~Brandstetter, M.~Hofmann, S.~Soff, C.~Spieles and H.~Weber, \textit{et al.}
%``Local thermodynamical equilibration in central Au + Au collisions at AGS,''
Phys. Lett. B \textbf{434}, 379-387 (1998)
doi:10.1016/S0370-2693(98)00624-8
[arXiv:nucl-th/9804008 [nucl-th]].
%66 citations counted in INSPIRE as of 03 Dec 2022

%\cite{Bravina:1999dh}
\bibitem{Bravina:1999dh}
L.~V.~Bravina, E.~E.~Zabrodin, M.~I.~Gorenstein, S.~A.~Bass, M.~Belkacem, M.~Bleicher, M.~Brandstetter, C.~Ernst, M.~Hofmann and L.~Neise, \textit{et al.}
%``Local equilibrium in heavy ion collisions: Microscopic model versus statistical model analysis,''
Phys. Rev. C \textbf{60}, 024904 (1999)
doi:10.1103/PhysRevC.60.024904
[arXiv:hep-ph/9906548 [hep-ph]].
%115 citations counted in INSPIRE as of 02 Feb 2023

%\cite{Huovinen:2002im}
\bibitem{Huovinen:2002im}
P.~Huovinen, M.~Belkacem, P.~J.~Ellis and J.~I.~Kapusta,
%``Dileptons and photons from coarse grained microscopic dynamics and hydrodynamics compared to experimental data,''
Phys. Rev. C \textbf{66}, 014903 (2002)
doi:10.1103/PhysRevC.66.014903
[arXiv:nucl-th/0203023 [nucl-th]].
%54 citations counted in INSPIRE as of 14 Dec 2022

%\cite{Endres:2013daa}
\bibitem{Endres:2013daa}
S.~Endres, H.~van Hees and M.~Bleicher,
%``Studies of Dilepton Production in Coarse-Grained Transport Dynamics,''
PoS \textbf{CPOD2013}, 052 (2013)
doi:10.22323/1.185.0052
%4 citations counted in INSPIRE as of 03 Dec 2022

%\cite{Endres:2014zua}
\bibitem{Endres:2014zua}
S.~Endres, H.~van Hees, J.~Weil and M.~Bleicher,
%``Coarse-graining approach for dilepton production at energies available at the CERN Super Proton Synchrotron,''
Phys. Rev. C \textbf{91}, no.5, 054911 (2015)
doi:10.1103/PhysRevC.91.054911
[arXiv:1412.1965 [nucl-th]].
%35 citations counted in INSPIRE as of 03 Dec 2022

%\cite{Endres:2015fna}
\bibitem{Endres:2015fna}
S.~Endres, H.~van Hees, J.~Weil and M.~Bleicher,
%``Dilepton production and reaction dynamics in heavy-ion collisions at SIS energies from coarse-grained transport simulations,''
Phys. Rev. C \textbf{92}, no.1, 014911 (2015)
doi:10.1103/PhysRevC.92.014911
[arXiv:1505.06131 [nucl-th]].
%60 citations counted in INSPIRE as of 02 Jan 2023

%\cite{Endres:2015egk}
\bibitem{Endres:2015egk}
S.~Endres, H.~van Hees and M.~Bleicher,
%``Photon and dilepton production at the Facility for Antiproton and Ion Research and beam-energy scan at the Relativistic Heavy-Ion Collider using coarse-grained microscopic transport simulations,''
Phys. Rev. C \textbf{93}, no.5, 054901 (2016)
doi:10.1103/PhysRevC.93.054901
[arXiv:1512.06549 [nucl-th]].
%33 citations counted in INSPIRE as of 07 Feb 2023

%\cite{Galatyuk:2015pkq}
\bibitem{Galatyuk:2015pkq}
T.~Galatyuk, P.~M.~Hohler, R.~Rapp, F.~Seck and J.~Stroth,
%``Thermal Dileptons from Coarse-Grained Transport as Fireball Probes at SIS Energies,''
Eur. Phys. J. A \textbf{52}, no.5, 131 (2016)
doi:10.1140/epja/i2016-16131-1
[arXiv:1512.08688 [nucl-th]].
%70 citations counted in INSPIRE as of 21 Feb 2023

%\cite{Endres:2016tkg}
\bibitem{Endres:2016tkg}
S.~Endres, H.~van Hees and M.~Bleicher,
%``Energy, centrality and momentum dependence of dielectron production at collider energies in a coarse-grained transport approach,''
Phys. Rev. C \textbf{94}, no.2, 024912 (2016)
doi:10.1103/PhysRevC.94.024912
[arXiv:1604.06415 [nucl-th]].
%15 citations counted in INSPIRE as of 29 Dec 2022

%\cite{Reichert:2020yhx}
\bibitem{Reichert:2020yhx}
T.~Reichert, G.~Inghirami and M.~Bleicher,
%``Probing chemical freeze-out criteria in relativistic nuclear collisions with coarse grained transport simulations,''
Eur. Phys. J. A \textbf{56}, no.10, 267 (2020)
doi:10.1140/epja/s10050-020-00273-y
[arXiv:2007.06440 [nucl-th]].
%12 citations counted in INSPIRE as of 05 Dec 2022

%\cite{Rapp:1999us}
\bibitem{Rapp:1999us}
R.~Rapp and J.~Wambach,
%``Low mass dileptons at the CERN SPS: Evidence for chiral restoration?,''
Eur. Phys. J. A \textbf{6}, 415-420 (1999)
doi:10.1007/s100500050364
[arXiv:hep-ph/9907502 [hep-ph]].
%265 citations counted in INSPIRE as of 12 Jan 2023

%\cite{vanHees:2007th}
\bibitem{vanHees:2007th}
H.~van Hees and R.~Rapp,
%``Dilepton Radiation at the CERN Super Proton Synchrotron,''
Nucl. Phys. A \textbf{806}, 339-387 (2008)
doi:10.1016/j.nuclphysa.2008.03.009
[arXiv:0711.3444 [hep-ph]].
%189 citations counted in INSPIRE as of 12 Jan 2023

%\cite{Rapp:2013nxa}
\bibitem{Rapp:2013nxa}
R.~Rapp,
%``Dilepton Spectroscopy of QCD Matter at Collider Energies,''
Adv. High Energy Phys. \textbf{2013}, 148253 (2013)
doi:10.1155/2013/148253
[arXiv:1304.2309 [hep-ph]].
%121 citations counted in INSPIRE as of 03 Feb 2023

%\cite{Chatterjee:2005de}
\bibitem{Chatterjee:2005de}
R.~Chatterjee, E.~S.~Frodermann, U.~W.~Heinz and D.~K.~Srivastava,
%``Elliptic flow of thermal photons in relativistic nuclear collisions,''
Phys. Rev. Lett. \textbf{96}, 202302 (2006)
doi:10.1103/PhysRevLett.96.202302
[arXiv:nucl-th/0511079 [nucl-th]].
%150 citations counted in INSPIRE as of 05 Jan 2023

%\cite{Chatterjee:2007xk}
\bibitem{Chatterjee:2007xk}
R.~Chatterjee, D.~K.~Srivastava, U.~W.~Heinz and C.~Gale,
%``Elliptic flow of thermal dileptons in relativistic nuclear collisions,''
Phys. Rev. C \textbf{75}, 054909 (2007)
doi:10.1103/PhysRevC.75.054909
[arXiv:nucl-th/0702039 [nucl-th]].
%65 citations counted in INSPIRE as of 29 Dec 2022

%\cite{Chatterjee:2008tp}
\bibitem{Chatterjee:2008tp}
R.~Chatterjee and D.~K.~Srivastava,
%``Elliptic flow of thermal photons and formation time of quark gluon plasma at RHIC,''
Phys. Rev. C \textbf{79}, 021901 (2009)
doi:10.1103/PhysRevC.79.021901
[arXiv:0809.0548 [nucl-th]].
%65 citations counted in INSPIRE as of 16 Feb 2023

%\cite{Deng:2010pq}
\bibitem{Deng:2010pq}
J.~Deng, Q.~Wang, N.~Xu and P.~Zhuang,
%``Dilepton flow and deconfinement phase transition in heavy ion collisions,''
Phys. Lett. B \textbf{701}, 581-586 (2011)
doi:10.1016/j.physletb.2011.06.027
[arXiv:1009.3091 [nucl-th]].
%23 citations counted in INSPIRE as of 29 Dec 2022

%\cite{vanHees:2011vb}
\bibitem{vanHees:2011vb}
H.~van Hees, C.~Gale and R.~Rapp,
%``Thermal Photons and Collective Flow at the Relativistic Heavy-Ion Collider,''
Phys. Rev. C \textbf{84}, 054906 (2011)
doi:10.1103/PhysRevC.84.054906
[arXiv:1108.2131 [hep-ph]].
%156 citations counted in INSPIRE as of 17 Feb 2023

%\cite{Mohanty:2011fp}
\bibitem{Mohanty:2011fp}
P.~Mohanty, V.~Roy, S.~Ghosh, S.~K.~Das, B.~Mohanty, S.~Sarkar, J.~e.~Alam and A.~K.~Chaudhuri,
%``Elliptic flow of thermal dileptons as a probe of QCD matter,''
Phys. Rev. C \textbf{85}, 031903 (2012)
doi:10.1103/PhysRevC.85.031903
[arXiv:1111.2159 [nucl-th]].
%21 citations counted in INSPIRE as of 29 Dec 2022

%\cite{Vujanovic:2013jpa}
\bibitem{Vujanovic:2013jpa}
G.~Vujanovic, C.~Young, B.~Schenke, R.~Rapp, S.~Jeon and C.~Gale,
%``Dilepton emission in high-energy heavy-ion collisions with viscous hydrodynamics,''
Phys. Rev. C \textbf{89}, no.3, 034904 (2014)
doi:10.1103/PhysRevC.89.034904
[arXiv:1312.0676 [nucl-th]].
%76 citations counted in INSPIRE as of 16 Dec 2022

%\cite{Vovchenko:2016ijt}
\bibitem{Vovchenko:2016ijt}
V.~Vovchenko, I.~A.~Karpenko, M.~I.~Gorenstein, L.~M.~Satarov, I.~N.~Mishustin, B.~K\"ampfer and H.~Stoecker,
%``Electromagnetic probes of a pure-glue initial state in nucleus-nucleus collisions at energies available at the CERN Large Hadron Collider,''
Phys. Rev. C \textbf{94}, no.2, 024906 (2016)
doi:10.1103/PhysRevC.94.024906
[arXiv:1604.06346 [nucl-th]].
%26 citations counted in INSPIRE as of 05 Jan 2023

%\cite{PHENIX:2011oxq}
\bibitem{PHENIX:2011oxq}
A.~Adare \textit{et al.} [PHENIX],
%``Observation of direct-photon collective flow in $\sqrt{s_{NN}}=200$ GeV Au+Au collisions,''
Phys. Rev. Lett. \textbf{109}, 122302 (2012)
doi:10.1103/PhysRevLett.109.122302
[arXiv:1105.4126 [nucl-ex]].
%281 citations counted in INSPIRE as of 27 Feb 2023

%\cite{Lohner:2012ct}
\bibitem{Lohner:2012ct}
D.~Lohner [ALICE],
%``Measurement of Direct-Photon Elliptic Flow in Pb-Pb Collisions at $\sqrt{s_{NN}} = 2.76$ TeV,''
J. Phys. Conf. Ser. \textbf{446}, 012028 (2013)
doi:10.1088/1742-6596/446/1/012028
[arXiv:1212.3995 [hep-ex]].
%143 citations counted in INSPIRE as of 02 Dec 2022

%\cite{ALICE:2018dti}
\bibitem{ALICE:2018dti}
S.~Acharya \textit{et al.} [ALICE],
%``Direct photon elliptic flow in Pb-Pb collisions at $\sqrt{s_{\rm NN}}$ = 2.76 TeV,''
Phys. Lett. B \textbf{789}, 308-322 (2019)
doi:10.1016/j.physletb.2018.11.039
[arXiv:1805.04403 [nucl-ex]].
%57 citations counted in INSPIRE as of 23 Feb 2023

%\cite{Fries:2003kq}
\bibitem{Fries:2003kq}
R.~J.~Fries, B.~Muller, C.~Nonaka and S.~A.~Bass,
%``Hadron production in heavy ion collisions: Fragmentation and recombination from a dense parton phase,''
Phys. Rev. C \textbf{68}, 044902 (2003)
doi:10.1103/PhysRevC.68.044902
[arXiv:nucl-th/0306027 [nucl-th]].
%691 citations counted in INSPIRE as of 24 Feb 2023

%\cite{LeFevre:2016vpp}
\bibitem{LeFevre:2016vpp}
A.~Le F\`evre, Y.~Leifels, C.~Hartnack and J.~Aichelin,
%``Origin of elliptic flow and its dependence on the equation of state in heavy ion reactions at intermediate energies,''
Phys. Rev. C \textbf{98}, no.3, 034901 (2018)
doi:10.1103/PhysRevC.98.034901
[arXiv:1611.07500 [nucl-th]].
%29 citations counted in INSPIRE as of 06 Feb 2023

%\cite{Fevre:2019xao}
\bibitem{Fevre:2019xao}
A.~L.~F\`evre, Y.~Leifels, J.~Aichelin and C.~Hartnack,
%``On the origin of the elliptic flow and its dependence on the Equation of State in heavy-ion reactions at intermediate energies,''
Nuovo Cim. C \textbf{41}, no.5, 180 (2019)
doi:10.1393/ncc/i2018-18180-x
%0 citations counted in INSPIRE as of 03 Dec 2022

%\cite{Gao:2022shr}
\bibitem{Gao:2022shr}
B.~Gao, Y.~Wang, Z.~Gao and Q.~Li,
%``Elliptic flow in heavy-ion collisions at intermediate energy: The role of impact parameter, mean field potential, and collision term,''
Phys. Lett. B \textbf{838}, 137685 (2023)
doi:10.1016/j.physletb.2023.137685
[arXiv:2210.08213 [nucl-th]].
%1 citations counted in INSPIRE as of 21 Feb 2023

\end{thebibliography}
\end{document}